\documentstyle[osa,manuscript,preprint]{revtex}  
\renewcommand{\baselinestretch}{0}
\begin{document}                
\preprint{SMUHEP 9708}
\title{\bf{Astrophysical Effects of $\nu \gamma \rightarrow \nu
\gamma \gamma$ and Its Crossed Processes}}
\author{Michael Harris, Jian Wang, Vigdor L. Teplitz}
\address{Department of Physics, Southern Methodist
University, Dallas, TX 75275}

\maketitle
\renewcommand{\baselinestretch}{0}
\begin{abstract}               
	Recently, Dicus and Repko computed  
$\nu \gamma \rightarrow \nu \gamma \gamma$ 
for energies below the threshold for $e^{+}e^{-}$
pair production. They found a cross section on the order of   
$10^{-52} \omega ^{\gamma }$ with
$ \gamma =10$, where $ \omega $ 
is the CMS energy of one
of the initial  particles in MeV.  Cross 
sections for the crossed processes are the same to factors of order one.  This 
note investigates the extent to which these processes could, if their result
extrapolates past $1$ MeV: affect supernova dynamics; cut off the energy 
distribution of very high energy cosmic photons and neutrinos; and possibly 
give rise to an observable gamma signal from scattering of neutrinos from one 
supernova by those of a second supernova close in space and time.  We also
estimate, from Supernova 1987A, that, in the region above a few MeV,  
$\gamma $ must
fall below $8.4$. 
\end{abstract}

\renewcommand{\baselinestretch}{0}
	The process $\nu \gamma \rightarrow \nu \gamma$ is quite small
(about $10^{-65}$ at $1 $ MeV) because  Yang's theorem [1] (two
photon decay of spin 1 particles of either parity forbidden) causes
it to vanish in the limit of zero momentum transfer (in lowest order).
The reaction $\nu \gamma \rightarrow \nu \gamma \gamma$          
and its crossed processes are not subject to this suppression [2,3] and 
hence are
potentially much more important to astrophysics.  The result of reference 3 is
\begin{eqnarray}
\sigma=\sigma_{0} ({\frac {\omega}{1 \mbox{\ }MeV}})^{\gamma} 
\mbox{\ \ \ \ \ \ }
\sigma _{0}=10^{-52}cm^{2} \mbox{\ \ \ \ \ \ }
\gamma=10  \mbox{\ \ \ \ \ \ }
\end{eqnarray}
for $\omega <0.5 $ MeV.  That paper speculates that $Eq.\mbox{\ }(1)$ 
 remains valid
for at least another order of magnitude in $\omega $.  However, due to 
the complexity
of the process, no evaluation of the cross section above the pair production
threshold has been done.  The purpose of this note is, in part, to investigate
the extent to which such evaluation might be motivated by astrophysical
applications -- particularly supernova
dynamics and attenuation of very high energy neutrinos and photons from
scattering off photon and neutrino backgrounds. We thus assume at first
that $Eq.\mbox{\ }(1)$ is
valid for $\omega $ in the region of tens of MeV's
\footnote{We note for comparison that, if the cross
section of Eq. (1) were limited to a small number of partial waves, the
partial wave unitarity limit would be exceeded for $\omega$ greater than
about $300$ MeV.}.

	The family of related processes
\begin{eqnarray} 
\nu \gamma \rightarrow \nu \gamma \gamma \mbox{\ \ (2a)\ \ \ \ \ }
\nu \bar{\nu } \rightarrow \gamma \gamma \gamma \mbox{\ \ (2b)\ \ \ \ \ }
\gamma \gamma \rightarrow \nu \bar{\nu } \gamma  \mbox{\ \ (2c)}
\end{eqnarray}
has several features of interest: all five particles are essentially massless
(CMS electron neutrino energies will be far above the $10$ eV extreme
limit on
its mass).  Thus the three processes have essentially the same cross section 
up to factors associated with spin and statistics.  Processes $(2a)$ and
$(2b)$ are potentially important contributions to supernova neutrino mean
free paths.  
They also contribute to ultra high energy photon scattering off cosmic
background neutrinos as
well as to ultra high energy scattering of neutrinos off various cosmic photon
backgrounds.  Process $(2c)$ is a possible energy loss mechanism for hot 
objects; it could be important in such contexts as the cooling of neutron
stars.

	We label the initial particles $q_{1}$ and $q_{2}$, the final ones
$p_{1}, p_{2}, p_{3}$, and the number of spin states $g(q_{i})$ and
$g(p_{i})$.
The rate
for interaction is then given by $[4]$
\begin{eqnarray}  
\frac{d  n_{12}}{d  t}=(2 \pi )^{-11} 2^{-5} g_{i} g_{f}
 \int\int \prod_{i=1}^{2} \frac{d^{3}  q_{i}}{q_{i}}
           \prod_{j=1}^{3} \frac{d^{3}  p_{i}}{p_{i}} 
           f(q_{1}) f(q_{2}) \prod_{k=1}^{3} (1\pm f(p_{k})) \nonumber \\
           \delta ^{4} [q_{1} +q_{2} -p_{1} -p_{2} -p_{3}] 
           \mid M\mid ^{2} 
\end{eqnarray} 
where
\[ g_{i}=g(q_{1}) g(q_{2}) \mbox{\ \ \ \ \ \ }
   g_{f}=g(p_{1}) g(p_{2}) g(p_{3})
\]
for processes  $(2a)$, $(2b)$ and $(2c)$ respectively, and 
the $f$'s are bose and
fermi distributions for $\gamma$ and $\nu$.  We will consider the 
possibility that
$T_{\gamma} \neq T_{\nu}$.  We ignore the factors of $g$ as well
as the angular
dependence of $M$.  From $Eq.\mbox{\ }(1)$ we have, after evaluating 
the CMS  phase space
integral
\begin{equation} 
\mid M\mid ^{2} = 2^{6} (2 \pi )^{3} \sigma_{0} \omega ^{\gamma }
\end{equation}
The rate for process (2a) then becomes
\begin{eqnarray} 
\frac{dn_{\nu \gamma}}{dt}=\frac{2 \sigma_{0}}{(2 \pi )^{8}}
      \int
          \frac{d^{3} q_{1}}{q_{1}} f_{F}(q_{1})
          \frac{d^{3} q_{2}}{q_{2}} f_{B}(q_{2})
          \frac{d^{3} p_{1}}{p_{1}} (1-f_{F}(p_{1}))
          \frac{d^{3} p_{2}}{p_{2}} (1+f_{B}(p_{2})) \nonumber \\
          \frac{d^{3}p_{3}}{p_{3}} (1+f_{B}(p_{3})) \omega ^{\gamma}   
          \delta ^{4} [q_{1} +q_{2} -p_{1} -p_{2} -p_{3}]
\end{eqnarray}
where ${\omega_{cm}}^{2}=q_{1}\cdot q_{2}/2$.  We eliminate the $q_{2}$ 
integral
with the $3-D$ delta function and write for the time component
\begin{equation} 
\delta ^{4} [q_{2} -(p_{1} +p_{2}+ p_{3} -q_{1})]= \nonumber \\
 q_{2} \delta \{ D p_{3} -[(p_{1} +p_{2})^{2}/2 -q_{1}(p_{1} +p_{2})
                            +\bf{q}_{1} \cdot (\bf{p}_{1} +\bf{p}_{2})]
               \}
\end{equation}
where
\[
D=q_{1} -p_{1} -p_{2}+(\bf{p}_{1} +\bf{p}_{2}) \cdot \hat{\bf{p}}_{3}
  -\bf{q}_{1} \cdot \hat{\bf{p}}_{3}
\]
with $\hat{\bf{p}}_{3}$ the unit vector in the direction of $\bf{p}_{3}$.
This gives
\begin{eqnarray} 
\frac{dn_{12}}{dt}=\frac{4 \sigma_{0}}{(2 \pi )^{6}}
                   \int 
                   d q_{1} d p_{1} d  p_{2}
                   d z_{1}d  z_{2}d  z_{3}
                   d \phi _{1} d  \phi _{2}   
                   q_{1} (p_{1} p_{2} p_{3})  
                   D^{-1}  \nonumber \\                  
                   f_{F}(q_{1}) f_{B}(q_{2})
                   \prod_{i=1}^{3}  (1\pm f(p_{i})) 
                   \Theta (q_{2}) {\omega _{cm}}^{\gamma }     
\end{eqnarray}
where we have taken $q_{1}$ along the z-axis and $p_{3}$ in the $x-z$ 
plane with no loss
of generality.   The expressions for processes (2b) and (2c) are the same
except for the fermi and bose factors.  From Eqs. (7) and (8), we can find the
mean free path, $\lambda _{1} $, for particle $1$ by
\begin{equation} 
\lambda _{1}= c n_{1}/(\frac{dn_{12}}{dt})
\end{equation} 
We evaluated the integral of $Eq.\mbox{\ }(7)$ for various cases with 
Monte Carlo
methods:  First, if $T_{\nu} = T_{\gamma} =T$ and $\mu_{\nu }=0$, one sees  
from dimensional analysis of Equations (7) and (8)   
that $\lambda _{1} $  
must go as $T^{-(\gamma +3)}$.  Thus we can write
\begin{equation} 
\lambda _{1}=\Lambda T^{-13} cm
\end{equation}

The Monte Carlo integration gives 
$\Lambda \sim 2.2 \times 10^{14}  cm MeV^{13}$ for $\gamma =10$. 
\  $T$ is measured
in MeV and the result is essentially the same for all three 
processes.\ $Table\mbox{\ } 1$ gives
results for several choices of $T$ and non-zero values of $\mu_{\nu }$ 
($\mu_{\bar{\nu }}=- \mu_{\nu }$).
The degeneracy parameter $\eta =\mu /T$ can be found
from the values given.
The important  
feature is that for $\omega \geq  5 $ MeV, the mean free path for a
neutrino
is less than the size, $10^{6}$ cm, of the collapsing core.
$Table \ 2$
gives results for several choices of $T_{\nu}$   and  $\mu_{\nu}$ holding
$T_{\gamma}$ fixed at
$5$ MeV.  Note from the tables that the effect of final state Bose-Einstein
enhancement for
$(2b)$ becomes dramatic for $T_{\nu}>T_{\gamma}$.  No results for $(2c)$
are given in
$Table\mbox{\  } 2$ because the variation is less than a factor of ten
over all the values
considered
\footnote{We have estimated relativistic plasma effects by evaluating the
integrals with integrands set to zero for photon energies less than the
plasma
frequency as given in, for example, E. Braaten and D. Segel, Phys. Rev.
D48,
1478 (1993).  The result is increase in mean free paths of less than 10
percent.}.

	We now turn to the implications of these results for supernova 
dynamics [5].  Since the mean free paths for the processes 
of $Eq.\mbox{\ }\ (2)$ are less than
the size of the supernova core ($10^{6}$ cm) for much of the parameter
space, it
would appear desirable that these processes be included in supernova codes. 
Some of the steps in which they might be particularly important include: 
$(i)$
The collapse phase.  During collapse, the effect of $\nu_{e} -e$ 
scattering is to
lower neutrino energies facilitating their escape and thereby to decrease the
electron fraction which, in turn, results in bounce from a smaller core and
hence a smaller shock to drive a larger mass [5,6].  The results of 
$Table \mbox{\ }2$ imply
that process $(2a)$ could play a similar role for $\omega$ over a few
MeV
and that 
therefore prompt shock mechanisms are only feasible if photon temperatures
during infall are kept below about $5$ MeV.
$(ii)$ 
Immediately after bounce.  At this time, the hot core has
temperatures on the order of $20-40$ MeV and neutrino energies range beyond 
100
MeV.  At these energies, the validity of $Eq.\mbox{\ } (1)$ is more
suspect, but
$Table\mbox{\ } 1$
shows that even a
much more gentle rise with temperature would have short mean free paths for 
the
processes of $Eq.\mbox{\ }\ (2)$. 
$(iii)$ 
The first second.  So long as  $T$  stays above a
few MeV, the processes of $Eq.\mbox{\ }\ (2)$ should remain important.  

Finally, we note that a bound can be placed on the cross section for
$(2a)$ from the observed
distribution of neutrinos from Supernova 1987A.  We can ask 
whether $Eq.\mbox{\ }  
(1)$ is consistent with the fitting parameters of roughly 4 seconds 
of neutrino 
radiation at a temperature of $4$ MeV [5].  A neutrino of about $1$ MeV or 
less 
leaves the supernova immediately without further scattering, so that we need 
the mean free path for (2a) with the final neutrino having energy less
than $1$ MeV to be greater than $10^{11}$ cm.  $Eq.\mbox{\ }\ (1)$ gives only
$5\times 10^{8}$ cm (when the
appropriate theta function is inserted into $Eq.\mbox{\ }\ (7)$).  
Replacing $\gamma =10$ in 
$Eq.(1)$ by $\gamma =8.4$ (for $E>1$ MeV) changes the result 
to $10^{11}$ cm [10].   

	A second area in which the processes of $Eq.\mbox{\ }\ (2)$ 
are potentially of
interest is that of long distance travel of extremely
energetic 
neutrinos and photons given the
existence of the cosmic  microwave gamma and neutrino backgrounds as well 
as IR
and optical backgrounds [7].  They will attenuate the high energy particle 
over
cosmological distances, $L$, for energies such that
\begin{equation}
\int \frac{dn_{B}}{dE_{B}}\sigma (E, E_{B}) dE_{B} L>1
\end{equation}
where $E$ and $E_{B}$ refer to the energetic and background particles
respectively. Suppose, for example, $\nu_{e}$ had a mass of $10$ eV.
$Eq.\ (2a)$  would give $L$ approximately
$10^{9}$ lys  at $E = 4\times 10^{16}$ eV,
corresponding to $\omega \sim 0.3$ GeV, were $Eq.\mbox{\ }\ (1)$ to be
trusted that far.  Using instead 
$\gamma =8.4$ would give $E = 4\times 10^{17}$ eV, corresponding 
to $\omega \sim 1$ GeV.  In $Table \mbox{\ }(3)$,
we list results for a few
other cases of interest.

	An amusing application of $(2b)$ is the case of two supernovae close
to each other in space and time.  Massive stars formed in giant molecular
clouds will be separated by distances on the order of $100$ light years
and $100$ years of each other and will have lifetimes on the order of 
a million
years.  The chances of two type II supernova explosions in one
cloud within
$100$ lys are perhaps non-negligible.  Gamma rays produced at the intersection
of the two expanding neutrinospheres would, like the radiation from the
supernovae themselves [8], be absorbed by the cloud,  but the signal from the
re-radiated IR could, in principle, be of interest.   The total energy in the
signal would be on the order of

\begin{equation}
  E_{T}=
  \int dE_{\bar{\nu}} dE_{\nu}
    \frac{dN(E_{\nu})}{dE_{\nu}} (E_{\nu}+\bar{E_{\nu}}) 
		 \frac{dn(E_{\bar{\nu}})}{dE_{\bar{\nu}}} \sigma c 
  \Delta t \sim 10^{29} ergs
\end{equation}
which, when spread over ten years is unfortunately far below the diffuse  
IR background.
For completeness, we also note that, in a cooling neutron
star, one can estimate from $Eq.\mbox{\ }\ (10)$ that $Eq.\mbox{\ }\ (2c)$ 
should go roughly as
$10^{31}T^{17}_9 erg/s$ which falls below the URCA process [9] ($5
\times 10^{39}T^8_9$) for temperatures under about an MeV.

	In conclusion, the processes of $Eq.\mbox{\ }\ (2)$ could be 
quite important in
astrophysics, particularly in supernova dynamics and long distance travel of
photons and neutrinos.  From Supernova 1987A, we can deduce 
that $Eq.\mbox{\ }\ (1)$
must be modified in the region  beyond an
MeV or so by, at least, reduction of
the
exponent $\gamma $ from $10$ to $8.4$.  Clearly a calculation of 
the processes of 
$Eq.\mbox{\ }\ (2)$ in the region above the pair production threshhold 
would be quite
useful.

\section*{Acknowledgments}

We have benefited from very helpful conversations and communications with 
Edward Baron, David Berley, David Branch, Duane Dicus, Mal Ruderman and 
 Ryszard Stroynowski.  We are also
grateful to  Eleana Tibuleac for  participating in the early stages of this
work.

\hspace{-0.5in}
\vspace{-1.0in}
\begin{table}
\caption{Mean free paths for the three processes of $Eq.\mbox{\ }(2)$ for
the values of temperature and chemical potential indicated.  Asterisks
indicate MFP's greater than a light year.}
\renewcommand{\arraystretch}{0.8}
\renewcommand{\tabcolsep}{0mm}
\begin{tabular}{|@{\ }llllllllr@{\ \ }|} \hline
\multicolumn{8}{|@{\ }l}{$\nu \gamma \rightarrow \nu \gamma
\gamma$\hspace{1in}Mean Free Path(cm)}  &$\mu_{\nu}$  \\ \hline
$3.0 \times 10^{14}$ &$2.4 \times 10^{10}$ &$1.5 \times 10^{5}$ 
&$2.0 \times 10^{1}$  &$1.2 \times 10^{-1}$ &$2.9 \times 10^{-3}$
&$1.8 \times 10^{-8}$ &$2.2 \times 10^{-12}$ &-50 \\ \hline
$2.1 \times 10^{14}$ &$2.8 \times 10^{10}$ &$1.6 \times 10^{5}$ 
&$2.3 \times 10^{1}$  &$1.2 \times 10^{-1}$ &$2.8 \times 10^{-3}$
&$1.9 \times 10^{-8}$ &$2.3 \times 10^{-12}$ & -30 \\ \hline
$2.1 \times 10^{14}$ &$2.6 \times 10^{10}$ &$1.9 \times 10^{5}$ 
&$2.3 \times 10^{1}$  &$1.2 \times 10^{-1}$ &$2.8 \times 10^{-3}$
&$1.8 \times 10^{-8}$ &$2.2 \times 10^{-12}$ &-20 \\ \hline
$2.0 \times 10^{14}$ &$2.4 \times 10^{10}$ &$1.9 \times 10^{5}$ 
&$2.3 \times 10^{1}$  &$1.1 \times 10^{-1}$ &$2.8 \times 10^{-3}$
&$1.9 \times 10^{-8}$ &$2.2 \times 10^{-12}$ & -10 \\ \hline
$2.2 \times 10^{14}$ &$2.7 \times 10^{10}$ &$1.8 \times 10^{5}$ 
&$2.2 \times 10^{1}$  &$1.1 \times 10^{-1}$ &$2.7 \times 10^{-3}$
&$1.8 \times 10^{-8}$ &$2.1 \times 10^{-12}$ & 0 \\ \hline
$8.7 \times 10^{12}$ &$1.0 \times 10^{10}$ &$1.5 \times 10^{5}$ 
&$2.0 \times 10^{1}$  &$1.1 \times 10^{-1}$ &$2.6 \times 10^{-3}$
&$1.8 \times 10^{-8}$ &$2.2 \times 10^{-12}$ & 10 \\ \hline
$3.4 \times 10^{13}$ &$4.4 \times 10^{9}$ &$7.4 \times 10^{4}$ 
&$1.7 \times 10^{1}$  &$9.5 \times 10^{-2}$ &$2.4 \times 10^{-3}$
&$1.8 \times 10^-{8}$ &$2.2 \times 10^{-12}$ & 25 \\ \hline
$1.3 \times 10^{13}$ &$1.0 \times 10^{9}$ &$3.0 \times 10^{4}$ 
&$9.0 \times 10^{0}$  &$7.4 \times 10^{-2}$ &$2.1 \times 10^{-3}$
&$1.7 \times 10^-{8}$ &$2.1 \times 10^{-12}$ &  50 \\ \hline
\multicolumn{9}{|@{\ }l|}{$\nu \bar{\nu} \rightarrow 
\gamma \gamma \gamma$} \\
\hline
$5.2 \times 10^{5}$ &$1.0 \times 10^{4}$   &$5.6 \times 10^{1}$ 
&$3.2 \times 10^{-1}$ &$8.7 \times 10^{-3}$  &$4.6 \times 10^{-4}$
&$1.3 \times 10^-{8}$ &$2.5 \times 10^{-12}$ & -50 \\ \hline
$2.2 \times 10^{7}$ &$5.4 \times 10^{5}$   &$1.1 \times 10^{3}$ 
&$2.3 \times 10^{0}$ &$3.1 \times 10^{-2}$  &$1.2 \times 10^{-3}$
&$1.9 \times 10^-{8}$ &$3.1 \times 10^{-12}$ & -30 \\ \hline
$5.3 \times 10^{8}$ &$8.9 \times 10^{6}$   &$7.6 \times 10^{3}$ 
&$6.0 \times 10^{0}$  &$5.8 \times 10^{-2}$  &$1.9 \times 10^{-3}$
&$2.2 \times 10^-{8}$ &$3.3 \times 10^{-12}$ & -20 \\ \hline
$6.9 \times 10^{10}$ &$3.9 \times 10^{8}$   &$4.9 \times 10^{4}$ 
&$1.6 \times 10^{1}$  &$1.1 \times 10^{-1}$  &$3.1 \times 10^{-3}$
&$2.8 \times 10^-{8}$ &$3.5 \times 10^{-12}$ &-10 \\ \hline
$4.0 \times 10^{14}$ &$4.9 \times 10^{10}$   &$3.2 \times 10^{5}$ 
&$4.0 \times 10^{1}$  &$2.0 \times 10^{-1}$  &$4.9 \times 10^{-3}$
&$3.3 \times 10^-{8}$ &$3.9 \times 10^{-12}$ &   0 \\ \hline
$2.5 \times 10^{17}$ &$1.6 \times 10^{12}$   &$1.7 \times 10^{6}$ 
&$9.5 \times 10^{1}$  &$3.7 \times 10^{-1}$  &$7.5 \times 10^{-3}$
&$3.9 \times 10^-{8}$ &$4.3 \times 10^{-12}$ &  10 \\ \hline
  ***  &$2.0 \times 10^{14}$   &$1.1 \times 10^{7}$ 
&$3.1 \times 10^{2}$  &$8.3 \times 10^{-1}$  &$1.5 \times 10^{-2}$
&$5.1 \times 10^-{8}$ &$5.0 \times 10^{-12}$ & 25\\ \hline
  ***  &  ***   &$2.2 \times 10^{8}$ 
&$1.4 \times 10^{3}$  &$2.8 \times 10^{0}$   &$3.9 \times 10^{-2}$
&$7.9 \times 10^-{8}$ &$6.3 \times 10^{-12}$ &  50\\ \hline
\multicolumn{9}{|@{\ }l|}{$\gamma 
\gamma \rightarrow \gamma \nu \bar{\nu}$} \\
\hline
  ***  &  ***  &$2.3 \times 10^{6}$ 
&$5.1 \times 10^{1}$  &$2.2 \times 10^{-1}$ &$4.7 \times 10^{-3}$
&$2.7 \times 10^-{8}$ &$3.2 \times 10^{-12}$ & -50 \\ \hline
  ***  &$6.0 \times 10^{12}$ &$5.3 \times 10^{5}$ 
&$4.0 \times 10^{1}$  &$1.9 \times 10^{-1}$ &$4.3 \times 10^{-3}$
&$2.8 \times 10^-{8}$ &$3.3 \times 10^{-12}$ & -30 \\ \hline
  ***  &$3.5 \times 10^{11}$ &$4.0 \times 10^{5}$ 
&$3.6 \times 10^{1}$  &$1.8 \times 10^{-1}$ &$4.1 \times 10^{-3}$
&$2.6 \times 10^-{8}$ &$3.2 \times 10^{-12}$ & -20 \\ \hline
$3.1 \times 10^{15}$ &$6.1 \times 10^{10}$ &$2.9 \times 10^{5}$ 
&$3.3\times 10^{1}$  &$1.7\times 10^{-1}$ &$4.0 \times 10^{-3}$
&$2.7 \times 10^-{8}$ &$3.2 \times 10^{-12}$ &-10 \\ \hline
$3.3 \times 10^{14}$ &$4.0 \times 10^{10}$ &$2.7 \times 10^{5}$ 
&$3.2\times 10^{1}$  &$1.7\times 10^{-1}$ &$3.9 \times 10^{-3}$
&$2.7 \times 10^-{8}$ &$3.2 \times 10^{-12}$ &  0 \\ \hline
$2.9 \times 10^{15}$ &$5.9 \times 10^{10}$ &$2.9\times 10^{5}$ 
&$3.3\times 10^{1}$  &$1.7\times 10^{-1}$ &$4.0 \times 10^{-3}$
&$2.7 \times 10^-{8}$ &$3.2 \times 10^{-12}$ &  10 \\ \hline
  ***  &$1.3 \times 10^{12}$ &$4.2\times 10^{5}$ 
&$3.8\times 10^{1}$  &$1.7\times 10^{-1}$ &$4.2 \times 10^{-3}$
&$2.7 \times 10^-{8}$ &$3.2 \times 10^{-12}$ &  25 \\ \hline
  ***  &  ***  &$2.2\times 10^{6}$ 
&$5.0\times 10^{1}$  &$2.2\times 10^{-1}$ &$4.7 \times 10^{-3}$
&$2.8 \times 10^-{8}$ &$3.3 \times 10^{-12}$ &  50 \\ \hline
$T=1$  &   2  &  5  &  10  &  15  &  20 &  50
&  100 MeV  & \\ \hline 
\end{tabular} 
\end{table}                                                                

\begin{table}
\caption{Mean free paths for the processes of $Eq.\mbox{\ }(2a)$ 
and $(2b)$ with the photon temperature held constant at $5MeV$.  
Process
$(2c)$ varies little and is
omitted.}
\renewcommand{\arraystretch}{0.8}
\renewcommand{\tabcolsep}{0mm}
\begin{tabular}{|@{\ }llllllllr@{\ \ }|} \hline
\multicolumn{8}{|@{\ }l}{$\nu \gamma \rightarrow \nu \gamma
\gamma$\hspace{1in}Mean Free Path(cm)}  &$\mu_{\nu}$  \\ \hline
$1.8 \times 10^{5}$   &$5.8 \times 10^{3}$   &$7.5 \times 10^{2}$ 
&$2.0 \times 10^{2}$  &$7.2 \times 10^{1}$   &$2.6 \times 10^{1}$
&$1.3 \times 10^{1}$  &$6.3 \times 10^{0}$ &0 \\ \hline
$1.5 \times 10^{5}$   &$5.4 \times 10^{3}$   &$7.6 \times 10^{2}$ 
&$1.8 \times 10^{2}$  &$6.9 \times 10^{1}$   &$2.4 \times 10^{1}$
&$1.2 \times 10^{1}$  &$7.3 \times 10^{0}$ &10 \\ \hline
$1.1 \times 10^{5}$   &$5.4 \times 10^{3}$   &$7.3 \times 10^{2}$ 
&$2.0 \times 10^{2}$  &$6.3 \times 10^{1}$   &$3.0 \times 10^{1}$
&$1.3 \times 10^{1}$  &$4.7 \times 10^{0}$ &20 \\ \hline
$7.4 \times 10^{4}$   &$4.7 \times 10^{3}$   &$7.9 \times 10^{2}$ 
&$1.6 \times 10^{2}$  &$6.3 \times 10^{1}$   &$2.5 \times 10^{1}$
&$1.3 \times 10^{1}$  &$9.9 \times 10^{0}$ &25 \\ \hline
$5.5 \times 10^{4}$   &$5.1 \times 10^{3}$   &$8.3 \times 10^{2}$ 
&$1.5 \times 10^{2}$  &$6.0 \times 10^{1}$   &$2.3 \times 10^{1}$
&$1.2 \times 10^{1}$  &$7.9 \times 10^{0}$ &30 \\ \hline
$4.7 \times 10^{4}$   &$4.3 \times 10^{3}$   &$7.5 \times 10^{2}$ 
&$1.9 \times 10^{2}$  &$6.1 \times 10^{1}$   &$2.3 \times 10^{1}$
&$1.3 \times 10^{1}$  &$5.8 \times 10^{0}$ &35 \\ \hline
$4.1 \times 10^{4}$   &$4.5 \times 10^{3}$   &$7.1 \times 10^{2}$ 
&$2.0 \times 10^{2}$  &$6.7 \times 10^{1}$   &$2.6 \times 10^{1}$
&$1.2 \times 10^{1}$  &$8.3 \times 10^{0}$ &40 \\ \hline
$3.8 \times 10^{4}$   &$4.0 \times 10^{3}$   &$7.0 \times 10^{2}$ 
&$1.9 \times 10^{2}$  &$6.1 \times 10^{1}$   &$2.4 \times 10^{1}$
&$1.7 \times 10^{1}$  &$7.9 \times 10^{0}$ &45 \\ \hline
\multicolumn{9}{|@{\ }l|}{$\nu \bar{\nu} \rightarrow 
\gamma \gamma \gamma$} \\
\hline
$3.2 \times 10^{5}$   &$4.7 \times 10^{1}$   &$2.5 \times 10^{-1}$ 
&$6.1 \times 10^{-3}$  &$3.3 \times 10^{-4}$ &$3.9 \times 10^{-5}$
&$7.0 \times 10^{-6}$  &$1.9 \times 10^{-6}$ &0 \\ \hline
$1.7 \times 10^{6}$   &$1.1 \times 10^{2}$   &$4.5 \times 10^{-1}$ 
&$9.2 \times 10^{-3}$  &$4.8 \times 10^{-4}$ &$5.2 \times 10^{-5}$
&$9.5 \times 10^{-6}$  &$2.4 \times 10^{-6}$ &10 \\ \hline
$7.3 \times 10^{6}$   &$2.5 \times 10^{2}$   &$8.0 \times 10^{-1}$ 
&$1.4 \times 10^{-2}$  &$6.7 \times 10^{-4}$ &$7.0 \times 10^{-5}$
&$1.2 \times 10^{-5}$  &$2.9 \times 10^{-6}$ &20 \\ \hline
$1.1 \times 10^{7}$   &$3.6 \times 10^{2}$   &$1.0 \times 10^{0}$ 
&$1.8 \times 10^{-2}$  &$7.9 \times 10^{-4}$ &$8.0 \times 10^{-5}$
&$1.4 \times 10^{-5}$  &$3.1 \times 10^{-6}$ &25 \\ \hline
$1.9 \times 10^{7}$   &$5.2 \times 10^{2}$   &$1.3 \times 10^{0}$ 
&$2.2 \times 10^{-2}$  &$9.4 \times 10^{-4}$ &$9.1 \times 10^{-5}$
&$1.5 \times 10^{-5}$  &$3.6 \times 10^{-6}$ &30 \\ \hline
$3.4 \times 10^{7}$   &$6.9 \times 10^{2}$   &$1.7 \times 10^{0}$ 
&$2.6 \times 10^{-2}$  &$1.1 \times 10^{-3}$ &$1.0 \times 10^{-4}$
&$1.7 \times 10^{-5}$  &$3.9 \times 10^{-6}$ &35 \\ \hline
$6.9 \times 10^{7}$   &$9.9 \times 10^{2}$   &$2.2 \times 10^{0}$ 
&$3.2 \times 10^{-2}$  &$1.3 \times 10^{-3}$ &$1.2 \times 10^{-4}$
&$1.9 \times 10^{-5}$  &$4.4 \times 10^{-6}$ &40 \\ \hline
$1.3 \times 10^{8}$   &$1.2 \times 10^{3}$   &$2.8 \times 10^{0}$ 
&$3.8 \times 10^{-2}$  &$1.5 \times 10^{-3}$ &$1.4 \times 10^{-4}$
&$2.2 \times 10^{-5}$  &$5.1 \times 10^{-6}$ &45 \\ \hline
$T=1$  &   2  &  5  &  10  &  15  &  20 &  50
&  100 MeV  & \\ \hline 
\end{tabular} 
\end{table}

\begin{table}
\caption{Energies at which mean free paths through cosmic 
backgrounds drop below $10^9$ light years, for both 
$\gamma =10$ and $\gamma =8.4$ in
$Eq. (1)$.}

\renewcommand{\tabcolsep}{0mm}
\begin{tabular}{
|@{\mbox{\ \ \ }}l@{\mbox{\ \ \ \ \ \ }}
|@{\mbox{\ \ \ }}l@{\mbox{\ \ \ \ \ \ \ \ \ \ \ \ \ }} 
|@{\mbox{\ \ \ }}l@{\mbox{\ \ \ \ \ \ }}
|@{\mbox{\ \ \ }}l@{\mbox{\ \ \ \ \ }}
|} 
\hline
         &                  
&\multicolumn{2}{@{}c|}{Energy at $\lambda=10^{9}$ lys}\\ 
Particle & Background  & $\gamma =10$ &
$\gamma =8.4$
\\ \hline
Neutrino & Photon($E=2 \times 10^{-4}eV$)
&$2 \times 10^{20}eV$  & $1 \times
10^{22}eV$ \\ \hline
Photon & Neutrino($m=10eV$)
&$4 \times 10^{16}eV$  & $4 \times
10^{17}eV$ \\ \hline
Neutrino & IR Photon($E=4 \times 10^{-3}-0.1eV$)
&$2 \times 10^{16}eV$  & $4 \times
10^{17}eV$ \\ \hline
Neutrino & Starlight Photon($E=0.1-10eV$)
&$1 \times 10^{15}eV$  & $5 \times
10^{16}eV$ \\ \hline
\end{tabular}
\end{table}


\begin{references}
\bibitem{1}C. N. Yang, Phys. Rev. D {\bf 77}, 242 (1950).
\bibitem{2}The process $\gamma \gamma \rightarrow \nu \bar{
\nu } \gamma $ \ seems to have been first investigated in 
        a prescient paper by  H. Y. Chiu and P. Morrison, 
	 Phys. Rev. Lett. {\bf 5}, 573 (1960).  More quantitative 
        work was done by E. P. Shabalin and H. V. Nguyen, J. Exptl.
        Phys.(U.S.S.R) {\bf 44},\  1003 (1963).       
\bibitem{3}D. A. Dicus and W. Repko, Phys. Rev. Lett., to be published.
\bibitem{4}E. W. Kolb, and M. S. Turner, {\it The Early Universe},  
  Addison-Wesley, New York, 1993.
\bibitem{5}H. Bethe, Rev. Mod. Phys. {\bf 62}, 801 (1990).
\bibitem{6}See, for example, E. A. Baron, J. Cooperstein, and S. Kahana, 
          Phys. Rev. Lett. {\bf 55}, 126 (1985). 
          E. A. Baron, J. Cooperstein, and S. Kahana,
          Nucl. Phys. A {\bf 440}, 744 (1985).
          E. A. Baron and J. Cooperstein,
          Astrophys. J. {\bf 353}, 597 (1990).
          S. W. Bruenn, Astrophys. J. Suppl. {\bf 58}, 771 (1985).
          E. S. Myra, S. A. Bludman, Y. Hoffman, I. Lichtenstadt, N. Sack, 
          and K. A. Van Riper, Astrophys. J. {\bf 318}, 744 (1987).
\bibitem{7}F. W. Stecker and O. C. De Jager, Space Sci. Rev., 1995. 
\bibitem{8}See J. M. Shull, Ap. J.{\bf 237}, 769 (1980) and   
           C. Wheeler, T. J. Mazurek and A. Sivaramakrishnan, Ap. J. 
           {\bf 237}, 781 (1980) for discussion of supernovae 
           inside molecular clouds.
\bibitem{9}S. L. Shapiro and S. A. Teukolsky, 
           {\it Black Holes, White Dwarfs
           and Neutron Stars: The Physics of Compact Objects}, 
           John Wiley, NY, 1983.
\bibitem{10}More precisely, Monte Carlo results are as follows: if we 
            ask that Eq. (1) be valid with $\gamma =10$ up to CMS 
            energy 
            $\omega_{1}$ with $5 \mbox{\ } MeV >\omega_{1} >1\mbox{\
            }MeV$, the time and temperature
            parameters from SN 1987A require that, for $\omega
            >\omega_{1}$, 
            the constant $\gamma$ be taken as
            $\gamma < 8.4-0.54(\omega_{1}-1)$.

\end{references}
\end{document}